\documentclass{article}

\usepackage{arxiv}

\usepackage{amsmath}

\usepackage[utf8]{inputenc} 
\usepackage[T1]{fontenc}    
\usepackage{hyperref}       
\usepackage{url}            
\usepackage{booktabs}       
\usepackage{amsfonts}       
\usepackage{nicefrac}       
\usepackage{microtype}      
\usepackage[capitalise]{cleveref}       
\usepackage{lipsum}         
\usepackage{graphicx}
\usepackage{natbib}
\usepackage{doi}

\usepackage{newtxtext}
\usepackage{amssymb}
\usepackage{setspace}
\usepackage{amsthm}
\usepackage{algorithm}
\usepackage{algpseudocode}
\usepackage{comment}
\usepackage[nolist]{acronym}
\usepackage{xcolor}
\usepackage{caption}
\usepackage{subcaption}
\usepackage{multirow}
\usepackage{booktabs}
\usepackage{tikz}

\usepackage{xr}
\newtheorem{theorem}{Theorem}
\newenvironment{sproof}{%
	\proof}{\endproof}

\title{Algorithmic Predation: Equilibrium Analysis in Dynamic Oligopolies with Smooth Market Sharing}

\newif\ifuniqueAffiliation

\ifuniqueAffiliation 
\author{ \href{https://orcid.org/0000-0002-5712-1706}{\includegraphics[scale=0.06]{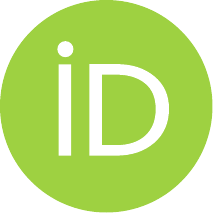}\hspace{1mm}Fabian R. Pieroth}\thanks{Use footnote for providing further
		information about author (webpage, alternative
		address)---\emph{not} for acknowledging funding agencies.} \\
	School of Computation, Information and Technology\\
	Technical University of Munich\\
	Munich, Germany \\
	\texttt{fabian.pieroth@tum.de} \\
	\And
	\href{https://orcid.org/0009-0005-7099-5685}{\includegraphics[scale=0.06]{orcid.pdf}\hspace{1mm}Ole Petersen} \\
	Department of Electrical Engineering\\
	Mount-Sheikh University\\
	Santa Narimana, Levand \\
	\texttt{stariate@ee.mount-sheikh.edu} \\
	\And
	\href{https://orcid.org/0000-0001-5491-2935}{\includegraphics[scale=0.06]{orcid.pdf}\hspace{1mm}Martin Bichler} \\
	School of Computation, Information and Technology\\
	Technical University of Munich\\
	Munich, Germany \\
	\texttt{bichler@cit.tum.de} \\
}
\else
\usepackage{authblk}

\newbox{\orcid}\sbox{\orcid}{\includegraphics[scale=0.06]{orcid.pdf}} 
\author[1, *]{%
	\href{https://orcid.org/0000-0002-5712-1706}{\usebox{\orcid}\hspace{1mm}Fabian R. Pieroth\thanks{\texttt{fabian.pieroth@tum.de}}}%
}
\author[1,2]{%
	\href{https://orcid.org/0009-0005-7099-5685}{\usebox{\orcid}\hspace{1mm}Ole Petersen\thanks{\texttt{ole.petersen@tum.de}}}%
}
\author[1]{%
	\href{https://orcid.org/0000-0001-5491-2935}{\usebox{\orcid}\hspace{1mm}Martin Bichler\thanks{\texttt{bichler@cit.tum.de}}}%
}
\affil[*]{These authors contributed equally to this work.}
\affil[1]{School of Computation, Information and Technology,
	Technical University of Munich,
	Munich, Germany}
\affil[2]{Listen Labs, San Francisco, CA, US}
\fi

\renewcommand{\shorttitle}{Algorithmic Predation}

\hypersetup{
pdftitle={Algorithmic Predation: Equilibrium Analysis in Dynamic Oligopolies with Smooth Market Sharing},
pdfsubject={q-bio.NC, q-bio.QM},
pdfauthor={Fabian R. Pieroth, Ole Petersen, Martin Bichler},
pdfkeywords={predatory prcing, oligopoly, dynamic game, equilibrium learning},
}

\begin{document}
\maketitle

\begin{abstract}
	Predatory pricing -- where a firm strategically lowers prices to undermine competitors -- is a contentious topic in dynamic oligopoly theory, with scholars debating practical relevance and the existence of predatory equilibria. Although finite-horizon dynamic models have long been proposed to capture the strategic intertemporal incentives of oligopolists, the existence and form of equilibrium strategies in settings that allow for firm exit (drop-outs following loss-making periods) have remained an open question. We focus on the seminal dynamic oligopoly model by Selten (1965) that introduces the subgame perfect equilibrium and analyzes smooth market sharing. Equilibrium can be derived analytically in models that do not allow for dropouts, but not in models that can lead to predatory pricing. In this paper, we leverage recent advances in deep reinforcement learning to compute and verify equilibria in finite-horizon dynamic oligopoly games.  
	Our experiments reveal two key findings: first, state-of-the-art deep reinforcement learning algorithms reliably converge to equilibrium in both perfect- and imperfect-information oligopoly models; second, when firms face asymmetric cost structures, the resulting equilibria exhibit predatory pricing behavior. These results demonstrate that predatory pricing can emerge as a rational equilibrium strategy across a broad variety of model settings. By providing equilibrium analysis of finite-horizon dynamic oligopoly models with drop-outs, our study answers a decade-old question and offers new insights for competition authorities and regulators.
\end{abstract}


\section{Introduction}
Predatory pricing is loosely defined as a firm's deliberate reduction of prices to levels that, while not necessarily below cost, are unsustainable for potential or existing competitors in the long run. 
In dynamic oligopoly competition, the strategic behavior associated with predatory pricing manifests as a dominant player systematically lowering prices to deter entry or push competitors out of the market~\citep{gatesDeterringPredationTelecommunications1995}. 

Antitrust laws, such as the Sherman Antitrust Act in the U.S. and Article 102 of the Treaty on the Functioning of the European Union (TFEU), address abusive practices like predatory pricing. For example, Article 102 of the TFEU prohibits a dominant firm from ''directly or indirectly imposing unfair purchase or selling prices.''  
However, whether predatory pricing is a concern in practice has long been controversial. 
\citet{dilorenzo1992myth} argues that while a firm might be able to successfully price other firms out of the market, there is no evidence to support the theory that the virtual monopoly could then raise prices. Also, courts have been skeptical of predatory pricing claims. For example, the U.S. Supreme Court has set high hurdles to antitrust claims based on predatory pricing theory~\citep{may1994brooke}. On the other hand, the US Department of Justice argues that predatory pricing is a real problem, courts are out of date, and too skeptical \citep{bolton1999predatory}.  
Predatory pricing has received renewed attention due to the presence of automated pricing agents in legal studies \citep{leslie2023predatory, cheng2023algorithmic}. Although a considerable body of scholarship aims to explain how algorithms can collude to fix prices \citep{bichler2014business}, almost no literature discusses anti-competitive behavior of algorithmic agents in the form of predatory pricing. 

In this article, we address two related problems:
First, can we expect algorithmic pricing agents in a dynamic or multi-stage oligopoly model to converge to an equilibrium? We focus on state-of-the-art deep reinforcement learning (DRL) algorithms as they constitute prime candidates for pricing agents in the field \citep{deng2024existence}. Convergence to an equilibrium is far from obvious, because we know that learning algorithms do not converge to equilibrium even in simple static games \citep{sanders2018prevalence}. Even less is known for multi-stage games. We draw on a recent approach to verify whether a strategy profile resulting from the interaction of DRL agents is a Nash equilibrium \citep{pieroth2025}. Second, we aim to understand in which environments we can expect a predatory equilibrium to emerge, and when this is not the case.

\subsection{Dynamic Oligopoly Competition}

Dynamic oligopoly models are well-suited to study predatory pricing, as firms interact over discrete time, repeatedly setting prices. Current choices influence future outcomes through mechanisms like demand inertia or strategic responses~\citep{milgromNewTheoriesPredatory1990}. Firms must be able to accumulate revenue over time, enabling recoupment of early losses, and must have the option to exit, drop out, or withdraw from the market to avoid further losses~\citep{telserCutthroatCompetitionLong1966}.

These interactions are often modeled as infinite-horizon stochastic games, where the Nash equilibrium (NE)~\citep{nash1950equilibrium} serves as the primary solution concept. Assuming complete information, these models can be solved using dynamic programming techniques, resulting in Markov Perfect Equilibria (MPE), a refinement of the NE \citep{maskin1988theory}. Previous literature showed the existence of MPEs displaying predatory behavior due to evasion of fixed costs or competitive advantage \citep{cabralLearningCurveMarket1994, besankoEconomicsPredationWhat, reyDynamicModelPredation2022}. 

Markov perfect equilibria (MPE) are not stable to small changes in payoffs and can shift discontinuously~\citep[518]{fudenberg1993learning}. Closed-form solutions are rare; instead, dynamic programming methods are used~\citep{pakes1992computing}, though convergence is not guaranteed and multiple MPEs may exist. By excluding history-dependent strategies, MPEs can miss key dynamics in settings with learning or private information.
These methods also face practical limits: they require discrete state and action spaces, full observability, and are difficult to extend to continuous or imperfect-information environments without high computational cost.

Deep reinforcement learning (DRL) methods such as PPO~\citep{DBLP:journals/corr/SchulmanWDRK17} can handle continuous action and state spaces, as well as imperfect information. \citet{pieroth2025} use DRL with self-play to learn candidate equilibria in multi-stage games, verifying convergence to a Nash equilibrium ex-post. These techniques constitute a breakthrough as they allow for equilibrium computation in finite-horizon, dynamic game-theoretical models. We build on this framework and present its first application for equilibrium analysis in dynamic oligopoly models.

\subsection{Contributions}

We study the dynamic oligopoly model of \citet{seltenSpieltheoretischeBehandlungOligopolmodells1965}, where $N$ firms produce a homogeneous good over a finite horizon $T$ (\cref{sec:dynamic_oligopoly_model}). Firms set prices from an interval, and market shares (demands) evolve based on price differences relative to the market average. This demand inertia -- capturing brand loyalty, switching costs, or network effects~\citep{besankoEconomicsPredationWhat} -- leads to \emph{smooth market sharing}, where small price cuts yield small market share gains. Unlike Bertrand competition~\citep{bertrandTheorieMathematiqueRichesse1883}, this avoids the paradox of prices being driven to marginal cost. Selten’s model yields more realistic pricing dynamics and has become influential for analyzing industries with limited price responsiveness, such as gasoline retail, banking, and telecommunications.

We compare two market models: firms either exit the market if unprofitable, like in contestable markets with a low barrier to entry, or persist despite losses. The model by \citet{seltenSpieltheoretischeBehandlungOligopolmodells1965} does not allow for {dropouts}, which are central to the analysis of predatory prices. The model with dropouts could lead to predatory pricing, where surviving firms capture increased market share and charge higher prices. However, if this can happen in equilibrium in a finite-horizon model is unkown. Although models with dropouts have been discussed \citep{bylkaDiscreteTimeDynamic2000}, this feature of the model is known to make the equilibrium analysis challenging. Each additional state grows the state space and numerical methods based on dynamic programming become very slow. 

We examine two information settings: in the perfect-information case, firms observe all demands after each round; in the imperfect-information case, they only observe current demand. This reflects real-world differences across markets, such as high transparency in gasoline or financial markets~\citep{gasoline,financial_markets} versus limited visibility in airlines or manufacturing~\citep{airline_demand_uncertainty}. While \citet{seltenSpieltheoretischeBehandlungOligopolmodells1965} solved the perfect-information case without dropouts, we provide an analytical characterization of the Nash equilibrium under imperfect information without dropouts. No analytical solution exists when dropouts are allowed.

We draw on the framework by \citet{pieroth2025} to compute a candidate approximate equilibrium strategy using Deep Reinforcement Learning (DRL), which is verified ex-post to confirm that it is indeed an approximate Nash equilibrium. These equilibrium guarantees are central to deep equlibrium learning and they allow us to verify Nash equilibria in finite-horizon games. The types of dynamic finite-horizon models in this paper could not be solved so far. 

This paper is the first application of deep equilibrium learning techniques in dynamic oligopolies leading to novel and policy-relevant insights. In particular, we show that predatory pricing arises as an equilibrium strategy in a wide variety of settings when firms can exit the market. That predatory pricing is possible in finite-horizon dynamic oligopoly competition models wtih continuous actions was an open question and we provide an affirmative answer to this policy-relevant question. 

The welfare analysis yields some counterintuitive results. While competition increases welfare in standard Bertrand oligopolies, this is not necessarily the case with smooth market sharing by \citet{seltenSpieltheoretischeBehandlungOligopolmodells1965}. Specifically, we find that predatory behavior leading to competitor exit can, under certain conditions, improve overall welfare. This occurs because the short-term aggressive pricing during predation often outweighs the subsequent higher prices during the recoupment phase. Additionally, exits typically involve less efficient firms, thereby raising market efficiency. These results challenge traditional antitrust perspectives, indicating that reductions in competition might somemathptmxyield welfare benefits, particularly when balanced against short recoupment windows and efficiency gains from market exits.

\section{Related work} \label{sec:related-work}

This section reviews related work on equilibrium analysis and learning dynamics in dynamic oligopoly models.
Dynamic oligopoly markets have been studied extensively in the literature \citep{fudenberg2013dynamic, gerpottCompetitivePricingOnline2022}. 

A foundational dynamic oligopoly model was introduced by \citet{seltenSpieltheoretischeBehandlungOligopolmodells1965}, who considered price competition with discrete time steps, finite horizon, complete information, and continuous demand. Selten explicitly characterized a deterministic subgame perfect equilibrium in a finite-horizon complete-information game, which was influential for subsequent analyses \citep{phlips1989dynamic, farrell1988dynamic, bayer2007network}. 
We extend his work and derive an equilibrium considering also imperfect information of firms. 

\citet{maskin1988theory} proposed an infinite-horizon model with alternating moves to study dynamic oligopolies, which focuses on long-run strategic considerations.
Several studies addressed predatory pricing within dynamic oligopolies in this framework~\citep{cabralLearningCurveMarket1994, besanko2014economics, reyDynamicModelPredation2022}. Despite their insights, these models often rely on strong assumptions, such as independent stage-wise demand, finite pay-off structures, or limited action spaces, limiting their ability to capture dynamic pricing behaviors. In contrast, our model incorporates interdependent demand and allows for continuous prices, enabling richer strategic patterns.


Finite-horizon models are arguably a good fit for the analysis of predatory pricing, as the strategic analysis of firms rarely considers an infinite horizon. They are less sensitive to discount factors or changes in the parameters of the game and an important complement to infinite-horizon and perfect-information models, for which numerical methods such as value function iteration have been available for a long time \citep{pakes1992computing}. However, solving finite-horizon models is challenging. \citet{bylkaDiscreteTimeDynamic2000} introduced dropout mechanisms, creating strategic discontinuities, which evaded equilibrium analysis so far. 
Furthermore, as the state space grows, numerical methods based on dynamic programming become slow quickly. Our approach, employing DRL, provides a way to find equilibrium even if the model allows dropouts, continuous actions, and states. 



Equilibrium learning offers an alternative numerical approach to finding equilibrium. It explores how equilibrium can emerge from agents that maximize their payoff while competing with each other~\citep{fudenbergTheoryLearningGames1999}. Almost the entire literature is focused on static, complete-information games. Unfortunately, learning dynamics does not necessarily converge to a Nash equilibrium~\citep{milionisImpossibilityTheoremGame2023, mazumdarPolicyGradientAlgorithmsHave2020, daskalakisLearningAlgorithmsNash2010}. 
Several recent studies have demonstrated the convergence of learning algorithms to equilibrium in static auction and oligopoly pricing models \citep{bichlerConvergenceLearningAlgorithms2023, serefahunbayUniquenessBayesianCoarse2024}. 

We build our study on a new methodology recently introduced by~\citet{pieroth2025}. They use deep \ac{RL} agents in self-play to compute candidate equilibrium profiles in multi-stage games with a finite horizon and continuous observations and actions. Importantly, they propose a verification algorithm that provides an upper bound on the computed candidate's distance to equilibrium. This enables an \textit{ex-post} verification of the learned strategies, offering guarantees even when there are none about convergence a priori.
We extend their work by studying dynamic oligopoly markets and computing novel approximate equilibrium strategies under various information structures and market rules. Additionally, we derive a novel equilibrium analytically, further contributing to the understanding of strategic behavior in these complex environments. This is the first work analyzing dynamic oligopoly models with this new equilibrium learning approach. 

\section{The Model}

We first outline the formal framework for \ac{MARL} and a suitable solution concept. Afterward, we introduce the dynamic oligopoly model considered.

\subsection{Partially observable Markov games}
We model the dynamic oligopoly as a \ac{POMG}, a generalization of a \ac{POMDP} for multiple agents \cite[Chapter 3.4]{marl-book}. 
Formally, a \ac{POMG} is a tuple $\langle \mathcal{S}, \mathcal{A}, \mathcal{T}, \mathcal{N}, \mathbf{r}, T, O, \mathbf{\Phi}, \mu \rangle$. Agents $i\in\mathcal{N}=\{1, \ldots, N\}$ collectively interact with an environment described by its state $s_t \in \mathcal{S}$ at time $t$. In each timestep, agents receive an observation $o_t^i=\Phi_i(s_t)$ with $o_t^i \in O_i$ and $O = \times_{i \in \mathcal{N}}O_i$. Subsequently, they choose an action $a_t^i \in \mathcal{A}_i$ according to their policy (or strategy) $\pi_i: O_i \rightarrow \Delta(\mathcal{A}_i)$, where $\mathcal{A}=\times_{i \in \mathcal{N}} \mathcal{A}_i$ and $\Delta(X)$ is the set of probability distributions over a set $X$. We denote the set of agent $i$'s policies by $\Sigma_i = \left\{\pi_i | \pi_i: O_i \rightarrow \Delta(\mathcal{A}_i) \right\}$. A policy is deterministic if it maps each observation $o_t^i$ on a specific action $a_t^i \in \mathcal{A}_i$. The environment transitions to a new state $s_{t+1} \sim \mathcal{T}(s_t, a_t^1, \ldots, a_t^N)$ and rewards each agent $i$ with $r_t^i = r_i(s_t, a_t^1, \ldots, a_t^N, s_{t+1})$. The goal of each agent is to maximize its expected cumulative reward or \textit{utility} $U_i(\pi_1, \ldots, \pi_N)=\mathbb{E}\left[\sum_{t=1}^{T}r_t^i\right]$. The game starts in an initial state $s_1\sim \mu$ and ends after $T$ timesteps.

We want to find an (approximate) \ac{NE}. A set of policies (also called \textit{strategy profile}) $\pi^*_{\mathcal{N}} \equiv \{\pi^*_1, \ldots, \pi^*_N\}$ is a $\varepsilon$-NE of a \ac{POMG} if and only if
\begin{align}
    \label{eq:ne_definition}
    \sup_{\pi_i \in \Sigma_i }U_i(\pi_i,\pi^*_{-i}) - U_i(\pi^*_\mathcal{N}) \leq \varepsilon \quad \forall i \in \mathcal{N},
\end{align}
where $\pi_{-i} \equiv \pi_{\mathcal{N}\setminus \{i\}}$. The strategy profile $\pi^*$ is denoted simply as a \ac{NE} if $\varepsilon=0$.

\subsection{Dynamic oligopoly model}
\label{sec:dynamic_oligopoly_model}
\begin{algorithm}[t]
    \begin{algorithmic}
        \Require
        \Statex Set of agents $\mathcal{N}=\{1,\dots,N\}$
        \Statex Number of rounds $T$
        \Statex For each agent $i$: initial demand $D_1^i$, unit production cost $c_i$, policy $\pi_i$, observation function $\Phi_i$
        \For{$t=1,2,\dots, T$}
        \For{$i \in \mathcal{N}$}
        \State $i$ observes $o_t^i = \Phi_i(s_t) = \Phi_i(t,D_t^1,\dots,D_t^N)$
        \State $i$ selects a price $p_t^i \sim \pi_i(o_t^i)$
        \State $i$ sells quantity $q_t^i=D_t^i - p_t^i$
        \State $i$ receives reward $r_t^i = (p_t^i - c_i)q_t^i$
        \EndFor
        \State Compute the average price as $\bar{p}_t = \frac{1}{N}\sum_{j \in \mathcal{N}} p_t^j$
        \For{$i \in \mathcal{N}$}
        \State Compute the price difference $\Delta p_t^i = p_t^i - \bar{p}_t$
        \State Transition demand to $D_{t+1}^i = D_t^i - \Delta p_t^i$
        \State Optionally, drop out $i$ if $D_{t+1}^i < c_i$ (see \cref{eq:dynamic_oligopoly_dropouts})
        \EndFor
        \EndFor
        \State Reward each agent $i$ with $U_i = \sum_{t=1}^T r_t^i$
    \end{algorithmic}
    \caption{Dynamic oligopoly game studied in this work.}
    \label{fig:dynamic_oligopoly:definition}
\end{algorithm}

We study an oligopoly model (see \cref{fig:dynamic_oligopoly:definition}) based on \citet{seltenSpieltheoretischeBehandlungOligopolmodells1965}, and incorporate a dropout mechanism inspired by \citet{bylkaDiscreteTimeDynamic2000}. Further, we introduce a novel imperfect information setting that considers uncertainty in real-world markets. The model consists of \(N\) firms producing a homogeneous good over a fixed time horizon \(T\). Each firm \(i\) has a constant unit production cost \(c_i\) and an initial demand \(D_1^i\).
$D_{t}^i$ is assumed to be the intercept of the inverse demand curve, that is, it represents the price at which the quantity demanded drops to zero. In each period \(t\), firms simultaneously set prices \(p_t^i\) from a continuous interval. Based on a linear demand model, firm \(i\) sells a quantity of \(D_t^i - p_t^i\) units, yielding a profit of \(r_t^i = (p_t^i - c_i)(D_t^i - p_t^i)\).
After all prices are set in period $t$, a below average price for firm $i$ attracts more customers, leading to increased demand \(D_{t+1}^i = D_t^i + \bar{p}_t - p_t^i\), where \(\bar{p}_t = \frac{1}{N}\sum_{j=1}^N p_t^j\) is the average price in period \(t\). The effect that not all customers immediately switch to the firm with the lowest price is demand inertia (\citet{seltenSpieltheoretischeBehandlungOligopolmodells1965} and can be due to switching costs or behavioral effects such as brand loyalty.

In the formulation as a \ac{POMG}, the state \(s_t\) includes the demand of each firm \(D_t^i\) and the current period \(t\). The action space \(\mathcal{A}_i=[c_i, p_{\text{max}}]\) comprises all possible prices \(p_t^i\) that firm $i$ can set, where the lower bound prevents selling at a loss and the upper bound $p_{\text{max}}$ is the monopolistic price. Agents \(\mathcal{N}\), the reward function \(\mathbf{r}\), transition function \(\mathcal{T}\), and the time horizon \(T\) align with the model description. 

We consider two different \textit{information settings}. The first is the fully observable case \(\Phi_i(s_t)=s_t=(t, D_t^1, D_t^2,\dots, D_t^N)\), where the firms observe the entire state, as in \citet{seltenSpieltheoretischeBehandlungOligopolmodells1965} and \citet{bylkaDiscreteTimeDynamic2000}. The second is the partially observable case, where firms only observe demand at the current time $t$, that is, $\Phi_i(s_t)=t$. This setting is relevant for markets where firms lack precise demand information, such as in online retail markets \citep{vandegeerDynamicPricingLearning2019} or ticket sales in the entertainment industry \citep{a06c4670-206d-3c08-987c-97337c3ffec5}. Such conditions are common in which firms protect their demand data and must infer their own demand from historical data \citep{vandegeerDynamicPricingLearning2019}.

A unique deterministic \ac{NE} of the form $p_i(D_t^1, \ldots, D_t^N,t)=\lambda_{1,t,i}+D_t^i \cdot \lambda_{2,t,i}$ is known for the case of complete observability~\citep{seltenSpieltheoretischeBehandlungOligopolmodells1965}.
To study predatory behavior, we extend Selten's model with a dropout mechanism inspired by \citet{bylkaDiscreteTimeDynamic2000}. However, since the number of customers of a firm is the \emph{area under the demand curve} rather than the demand itself, we preserve the total area under the demand curve after dropouts, yielding the following demand update:
\begin{align}
    \label{eq:dynamic_oligopoly_dropouts}
    \tilde{D}_{t+1}^i & =D_t^i+\bar{p_t}-p_t^i \:                                                                                      &     \\
    J_t &= \{i \in \mathcal{N} | \tilde{D}_{t+1}^i < c_i \}\\
    \bar{D}_{t+1}^i   & = \begin{cases}
                              \tilde{D}_{t+1}^i \text{ if } i \in \mathcal{N} \setminus J_t \\
                              0 \text{ otherwise}
                          \end{cases} \:                                                    &                                     \\
    D_{t+1}^i         & = \sqrt{ \left(\bar{D}_{t+1}^i\right)^2 + \frac{\bar{D}_{t+1}^i}{\sum_{k \in \mathcal{N} \setminus J_t} \bar{D}_{t+1}^k} \cdot \sum_{j \in J_t} \left(\tilde{D}_{t+1}^j \right)^2 } \: & 
\end{align}

Increasing prices in a stage increases short-term profits at the cost of losing market share in subsequent periods due to demand inertia. Capturing an early market share advantage thus yields significant benefits over multiple future periods. These competing incentives typically result in aggressive pricing early on, followed by price increases toward the end of the finite horizon. 

Introducing the possibility that firms may permanently exit the market amplifies these competitive dynamics. Specifically, the irreversible threat of market exit leads to even more aggressive pricing initially, as firms aim to survive and eliminate competitors. Once rivals are pushed out, the remaining firms gain additional market share, further enabling price increases in later rounds. The combination of demand inertia, stage-wise monopoly incentives, and the credible threat of permanent market exit makes Selten’s extended framework particularly suitable for studying predatory behavior. Moreover, the complexity introduced by a finite time horizon, interdependent demands, and dropout mechanisms has prevented analytical equilibrium analysis so far.


\section{Analytical equilibrium analysis of the dynamic model without demand observation}

We derive a deterministic \ac{NE} in the partially observable dynamic oligopoly without dropouts, complementing the one derived by \citet{seltenSpieltheoretischeBehandlungOligopolmodells1965} for fully observable markets:
\begin{theorem}\label{thm:ne_no_demand_observation}
    Consider a dynamic oligopoly model with \(N\) firms, unit production costs \(c_i\), initial demand \(D_1^i\), and time horizon \(T\). The model assumes no demand observation, i.e., \(\Phi_i(s_t)=t\), and no dropouts. 
    Then, any solution to the following system of equations constitutes a deterministic \ac{NE}:
    \begin{align}
        \label{eq:ne_no_demand_observation:optimality}
        &(D_t^i-2p_t^i+c_i)-\sum_{\tau=t+1}^{T} \left((p_\tau^i-c_i) \cdot \frac{N-1}{N} \right) \notag\\
        &\hspace{3.6cm} =0\quad \forall i \in \mathcal{N}, 1 \leq t \leq T \\
        \label{eq:ne_no_demand_observation:demand_update}
        &D_{t+1}^i =D_t^i-p_t^i+ \frac{1}{N}\sum_{j \in \mathcal{N}} p_t^j \quad \forall i \in \mathcal{N}, 1 \leq t < T,
    \end{align}
    where the constraints are $D_t^i\geq 0$ and $c_i \leq p_t^i < p_{\text{max}}$ for $1 \leq t \leq T$ and $i \in \mathcal{N}$.
\end{theorem}

\begin{sproof}
    \Cref{eq:ne_no_demand_observation:demand_update} follows from the demand update step. We then observe that the rewards are continuously differentiable. \Cref{eq:ne_no_demand_observation:optimality} is derived from the first-order condition $\frac{d U_i}{d p_t^i}=~0$, which gives us a necessary condition for a NE. We further check the second-order condition for a solution of the first-order condition, giving us a sufficient condition for a NE. 
\end{sproof}

\section{Learning in Markov Games}
\label{sec:learning_in_dynamic_oligopoly_models}

Classical \ac{RL} algorithms solve \acp{MDP}, where a single agent interacts with the environment. A straightforward approach to extend these algorithms to \acp{POMG} is \textit{self-play}. Here, independent instances of a single-agent \ac{RL} algorithm are employed for each agent, all interacting within the same environment \cite[Chapter 9.3.2]{marl-book}.
We consider \textit{policy gradient algorithms}, where each agent's policy $\pi_{\theta_i}(o_i)=\pi(\cdot | o,\theta_i)$ is parameterized by a neural network with parameters $\theta_i$. For continuous action spaces, the network outputs parameters of a continuous distribution, e.g., a normal or beta distribution. Parameters are updated simultaneously for all agents in each iteration according to:

\begin{align}
    \label{eq:pg}
    \theta_i \leftarrow \theta_i + \alpha \nabla_{\theta_i} U_i(\pi_{\theta_i},\{\pi_{\theta_j}\}_{j\in {\mathcal{N}\setminus \{i\}}})
\end{align}

The policy gradient $\nabla_{\theta_i} U_i$ can be estimated from a batch of game trajectories using the \textsc{Reinforce} algorithm or its variants \cite{SuttonAndBarto2018}, such as \ac{PPO}. In this work, we use both \textsc{Reinforce} and \ac{PPO} as implemented by \citet{raffinStableBaselines3ReliableReinforcement2021}.
After training, a pure strategy is extracted by selecting the most likely action.

\subsection{Measuring closeness to equilibrium} \label{sec:metrics}

We assess convergence to approximate \acp{NE} with a novel verification algorithm for multi-stage games with continuous states and actions introduced by \citet{pieroth2025}.
Given the learned strategy profile $\pi_\mathcal{N}$, it estimates the best-response utility $\sup_{\pi_i \in \Sigma_i} U_i(\pi_i,\pi_{-i})$ by discretizing the action- and observation spaces of agent $i$ and building up a game tree from the view of a single agent. For a given discretization $K\in \mathbb{N}$, it estimates the best-response utility by searching over a finite set of step functions $\Sigma_i^{K}$. Given large enough $K$, one has $\sup_{\pi_i \in \Sigma_i^{K}} U_i(\pi_i,\pi_{-i}) \approx \sup_{\pi_i \in \Sigma_i} U_i(\pi_i,\pi_{-i})$. We define the \textit{brute-force utility loss} for each agent $i \in \cal{N}$ as
\begin{align}
    \label{eq:bf_utility_loss}
    \mathcal{L}_{\text{bf},i}= \sup_{\pi_i \in \Sigma_i^{K}} U_i(\pi_i,\pi_{-i}) - U_i(\pi_i,\pi_{-i}).
\end{align}

The size of the game tree to build for this loss at a discretization $K$ scales exponentially in $T$, limiting the analysis to $T=4$ for $K=32$. Further, the brute-force verifier can only verify whether a given strategy profile is close to a \ac{NE} but not compute an approximate \ac{NE} itself.

Since the interpretation of the utility loss depends on the utility scale, we also report the \textit{normalized brute-force utility loss} $\mathcal{L}_{\text{bf,norm},i}= \mathcal{L}_{\text{bf},i}/ \max_{\pi_i \in \Sigma_i^{K}} U_i(\pi_i,\pi_{-i})$

\subsection{Measuring predatory behavior and its effects}

\citet{ordoverEconomicDefinitionPredation1981} characterize predatory pricing as a deliberate sacrifice of profits relative to a feasible, less aggressive action, followed by a recoupment of those losses once competitors exit the market. We develop a metric to measure the predatory incentive for each agent $i$ under strategy profile $\pi$, following that definition, employing the known analytical equilibrium strategies without dropouts as a baseline.

Denote by $\tau_i=\min \{t: \text{an opponent drops out}\}$ the first period in which an opponent exits. Let $r^{i, \text{equ}}_t$ represent agent $i$'s reward at time step $t$ when all agents follow the equilibrium strategy without dropouts for the whole game, and $r^{i, \pi}_t$ the corresponding reward under strategy profile $\pi$. Then, the predatory incentive for agent $i$ is defined as
\begin{align} \label{equ:predatory-incentives}
    \text{PI}_i(\pi) := \underbrace{- \sum_{t < \tau_i} \max\{0, r^{i, \text{equ}}_t - r^{i, \pi}_t \}}_{\text{sacrifice}} + \underbrace{\sum_{t \geq \tau} \max \{0, r^{i, \pi}_t - r^{i, \text{equ}}_t \}}_{\text{recoupment}}.
\end{align}
The first sum captures profit sacrificed before the rival’s exit, while the second sum measures subsequent recoupment gains. The use of the maximum operator ensures that only deliberate sacrifices and corresponding recoupment gains count toward the predatory incentive. If no opponent exists, we set $\text{PI}_i(\pi)=0$.
A strictly positive predatory incentive ($\text{PI}_i(\pi)>0$) indicates that agent $i$'s strategy, which induces an opponent's market exit, is ex-ante profitable relative to the non-exclusionary equilibrium benchmark. Conversely, a non-positive value ($\text{PI}_i(\pi)\leq 0$) implies that the observed pricing path lacks exclusionary justification.

To quantify the welfare implications of predatory pricing, we calculate total welfare of a strategy profile $\pi$ as the sum of consumer surplus and producer surplus over all periods: $W^\pi = \sum_{t=1}^T \left( \text{CS}_t^\pi + \text{PS}_t^\pi \right)$ \cite[p.\ 24]{belleflammeIndustrialOrganizationMarkets2010}. The producer surplus $\text{PS}_t^\pi =\sum_{i \in \mathcal{N}} r^{i, \pi}_t$ is the sum of all rewards.
The consumer surplus $\text{CS}_t^\pi := \sum_{i \in \mathcal{N}} (D_{t}^{i} - p_{t}^{i})q_{t}^{i} = \sum_{i \in \mathcal{N}}(D_{t}^{i} - p_{t}^{i})^2$ is the consumer's willingness to pay minus the price, following a linear demand model.

We measure welfare harm from predatory pricing by comparing welfare levels under dropout-enabled scenarios to the welfare in the corresponding analytical equilibrium without dropouts $\pi^*$, reporting the welfare difference $\Delta W^\pi := W^\pi - W^{\pi^*}$.

\section{Numerical equilibrium analysis experiments}
\label{sec:experimental_design}

In our numerical experiments, we conduct an equilibrium analysis of the introduced oligopolistic market to address three central questions: First, does predatory behavior emerge as a rational equilibrium strategy when firms can exit the market, and is it more profitable for the predator than the analytical equilibrium without dropouts? Second, how does predation affect consumer and producer welfare? And third, how sensitive are these outcomes to the information structure, specifically whether firms fully observe rivals' demand or operate under partial observability?


\subsection{Experimental design}

We consider the dynamic oligopoly from \cref{sec:dynamic_oligopoly_model} with $N=3$ agents, an initial demand of $D_1^i = 1$ for all $i \in \mathcal{N}$, and a time horizon of $T = 4$ stages. We evaluate brute-force utility loss, predatory incentives, and welfare differences across all combinations of the independent variables: \emph{information setting} (fully vs. partially observable), \emph{learning algorithm} (\ac{PPO} vs. \textsc{Reinforce}), and \emph{production costs}. For the latter, we examine asymmetries by fixing $c_1 = c_2 = 0.8$ and varying $c_0$ over $[0.42, 0.95]$ in 60 equidistant steps, yielding cost vectors $\mathbf{c} = [c_0, 0.8, 0.8]$.

We use a beta distribution for the action distribution, as suggested by \citep{petrazzini2021proximalpolicyoptimizationcontinuous}, with a fully connected network (3 linear layers, 64 units, SeLu activation) for all agents and algorithms. Each algorithm runs for $1,000$ iterations with $20,000$ trajectories per iteration at a learning rate of $8.57 \cdot 10^{-4}$ for PPO and $2.864 \cdot 10^{-4}$ for \textsc{Reinforce}. To improve accuracy, we divide the learning rate by eight for PPO and by two for \textsc{Reinforce} every $250$ iterations.

Training via self-play requires approximately $10$ minutes per run for \ac{PPO} and $6$ minutes for \textsc{Reinforce} on our hardware (GeForce RTX 2080 Ti, 12 Gb RAM). To cover the experimental design, we conduct $1,200$ training runs ($5$ seeds × $2$ information settings × $60$ production costs × $2$ algorithms), which can run in parallel.

\subsection{Results}

We now present the results of our equilibrium analysis. After a convergence analysis, we examine the emergence of distinct market regimes and predatory pricing behavior, followed by an evaluation of their welfare implications and sensitivity to the information structure.


\begin{figure*}[ht]
    \centering
    \includegraphics[width=0.5\columnwidth]{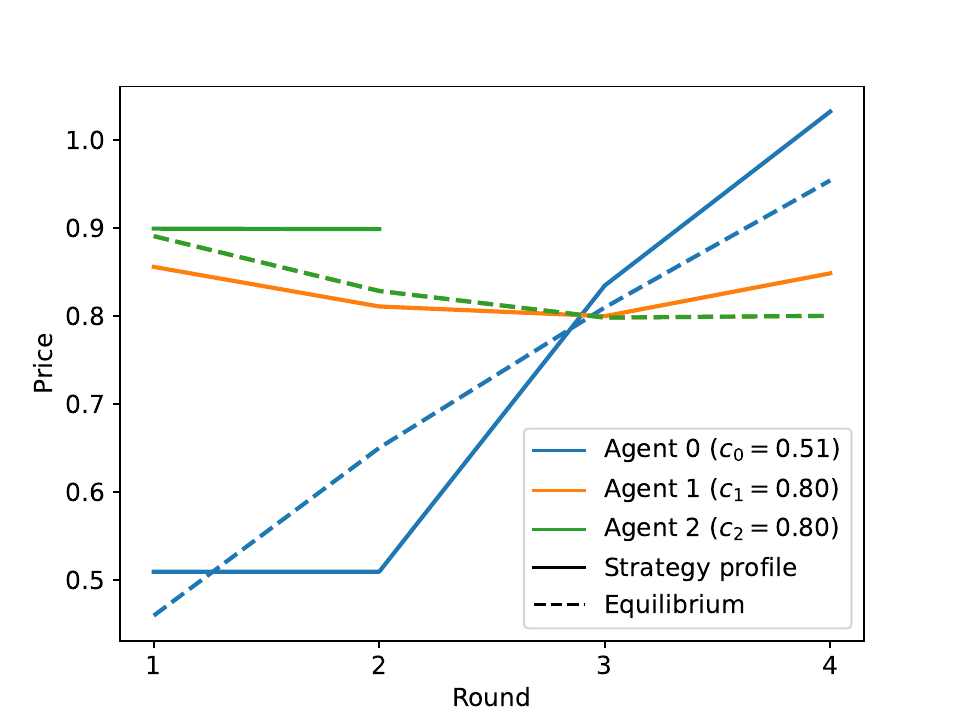}
    \caption{Strategy profile learned by \ac{PPO} in the partially observable case with dropouts for specific cost scenario $c_0=0.51$ and $c_1=c_2=0.8$. Recall that with partial observability, a deterministic probabilistic strategy is fully characterized by $T$ prices. If an agent drops out in a round, the graph stops at that round.}
    \label{fig:strategy_profiles_large_advantage}
\end{figure*}

\textbf{Equilibrium convergence and market regimes}: Table~\ref{tab:empirical_results:extreme-values-losses} shows that both PPO and \textsc{Reinforce} reliably converge to approximate equilibria with $\varepsilon \leq 0.032$ for all configurations studied. Therefore, we can confidently consider the following analyses as equilibrium analyses.

\begin{figure*}[ht]
    \centering
    \begin{subfigure}[t]{0.32\textwidth}
        \centering
        \includegraphics[width=\textwidth]{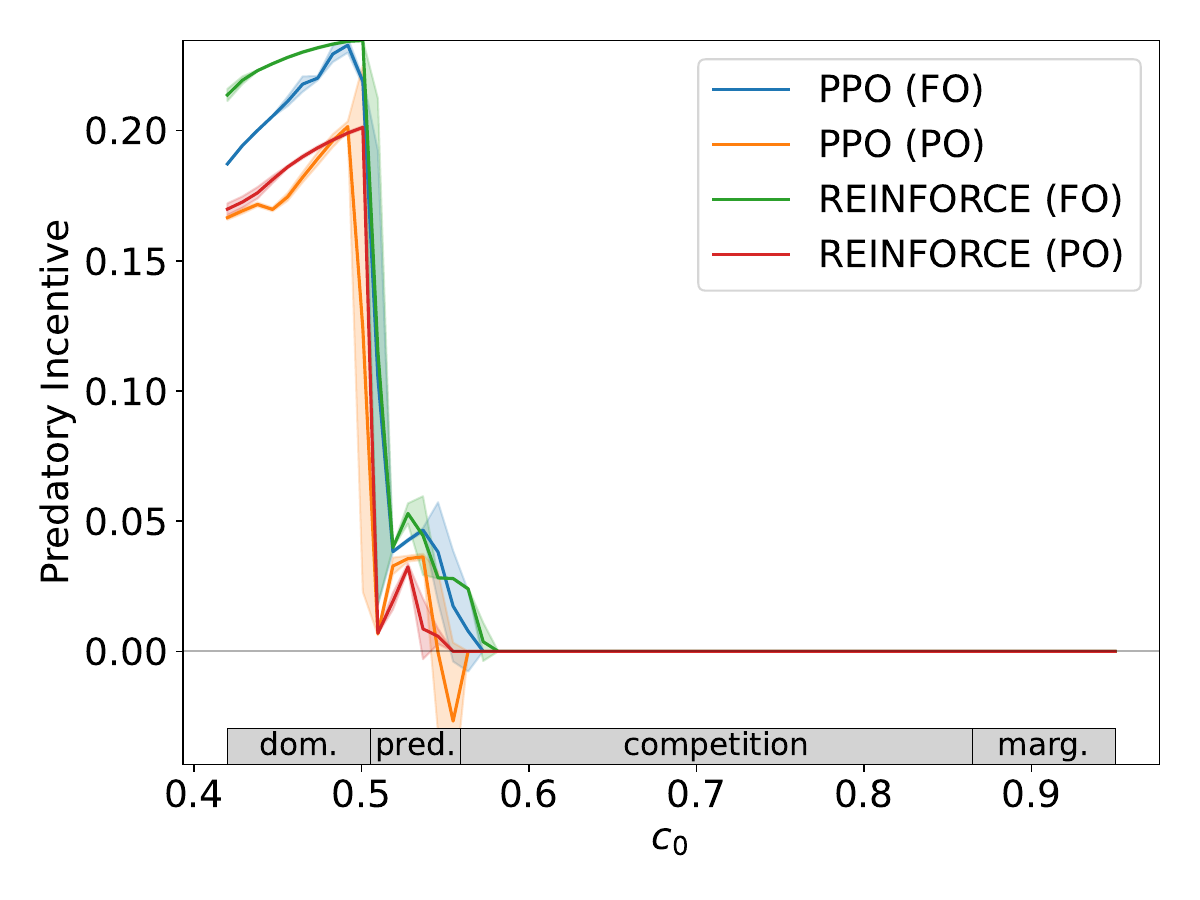}
        \caption{Agent $0$}
        \label{fig:predatory_incentives_agent_0}
    \end{subfigure}
    \hfill
    \begin{subfigure}[t]{0.32\textwidth}
        \centering
        \includegraphics[width=\textwidth]{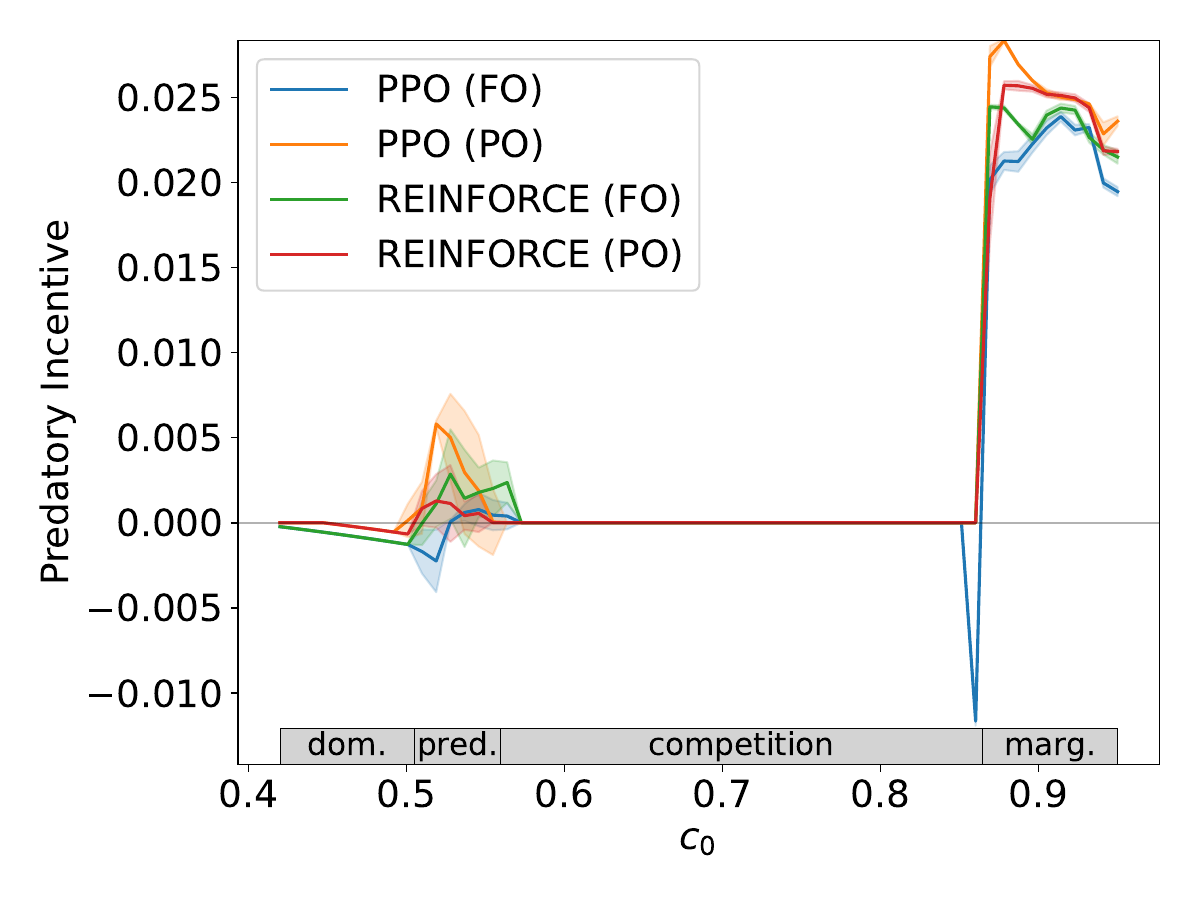}
        \caption{Agent $1$}
        \label{fig:predatory_incentives_agent_1}
    \end{subfigure}
    \hfill
    \begin{subfigure}[t]{0.32\textwidth}
        \centering
        \includegraphics[width=\textwidth]{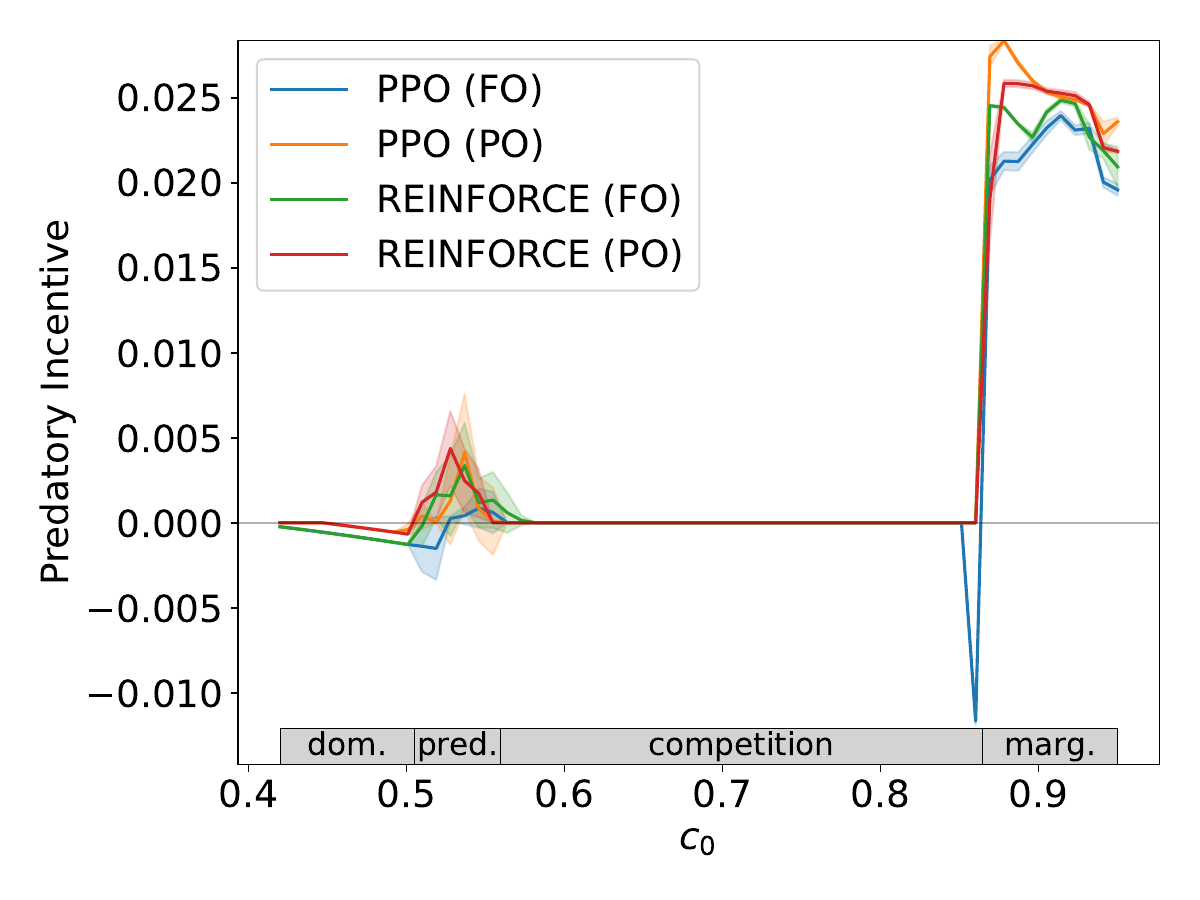}
        \caption{Agent $2$}
        \label{fig:predatory_incentives_agent_2}
    \end{subfigure}
    \hfill
    \caption{The predatory incentives $PI_i(\pi)$ for agents $i \in \{1, 2, 3\}$ and learned strategy profiles $\pi$ over the different costs $c_0$, information structures, and algorithms. The bold line represents the mean, and the colored shaded area represents the standard deviation over five seeds. The bottom bar indicates the regime, determined by a majority vote over all algorithms, information settings, and random seeds.}
    \label{fig:predatory_incentives}
\end{figure*}

\begin{figure*}[ht]
    \centering
    \begin{subfigure}[t]{0.32\textwidth}
        \centering
        \includegraphics[width=\textwidth]{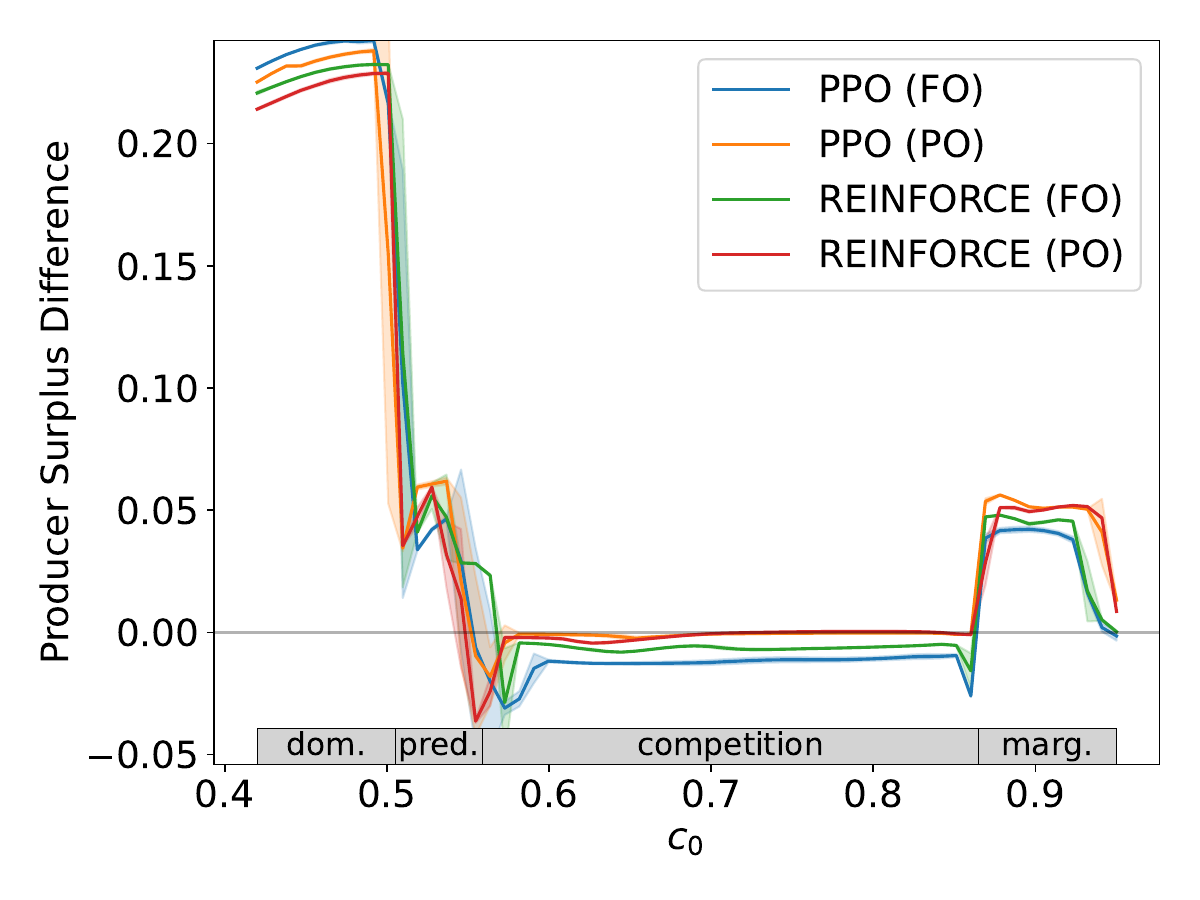}
        \caption{Producer surplus difference $\Delta PS^{\pi}$}
        \label{fig:producer-welfare-difference}
    \end{subfigure}
    \hfill
    \begin{subfigure}[t]{0.32\textwidth}
        \centering
        \includegraphics[width=\textwidth]{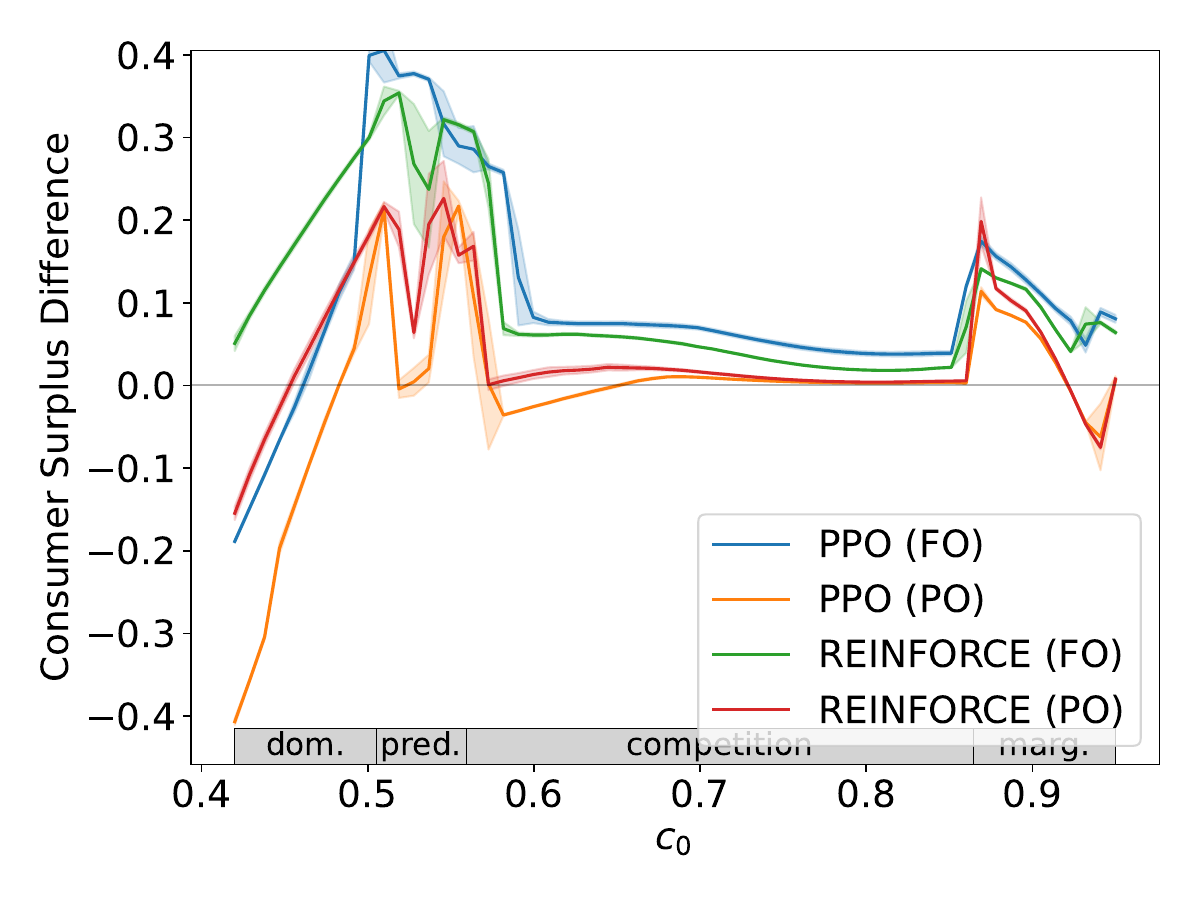}
        \caption{Consumer surplus difference $\Delta CS^{\pi}$}
        \label{fig:consumer-welfare-difference}
    \end{subfigure}
    \hfill
    \begin{subfigure}[t]{0.32\textwidth}
        \centering
        \includegraphics[width=\textwidth]{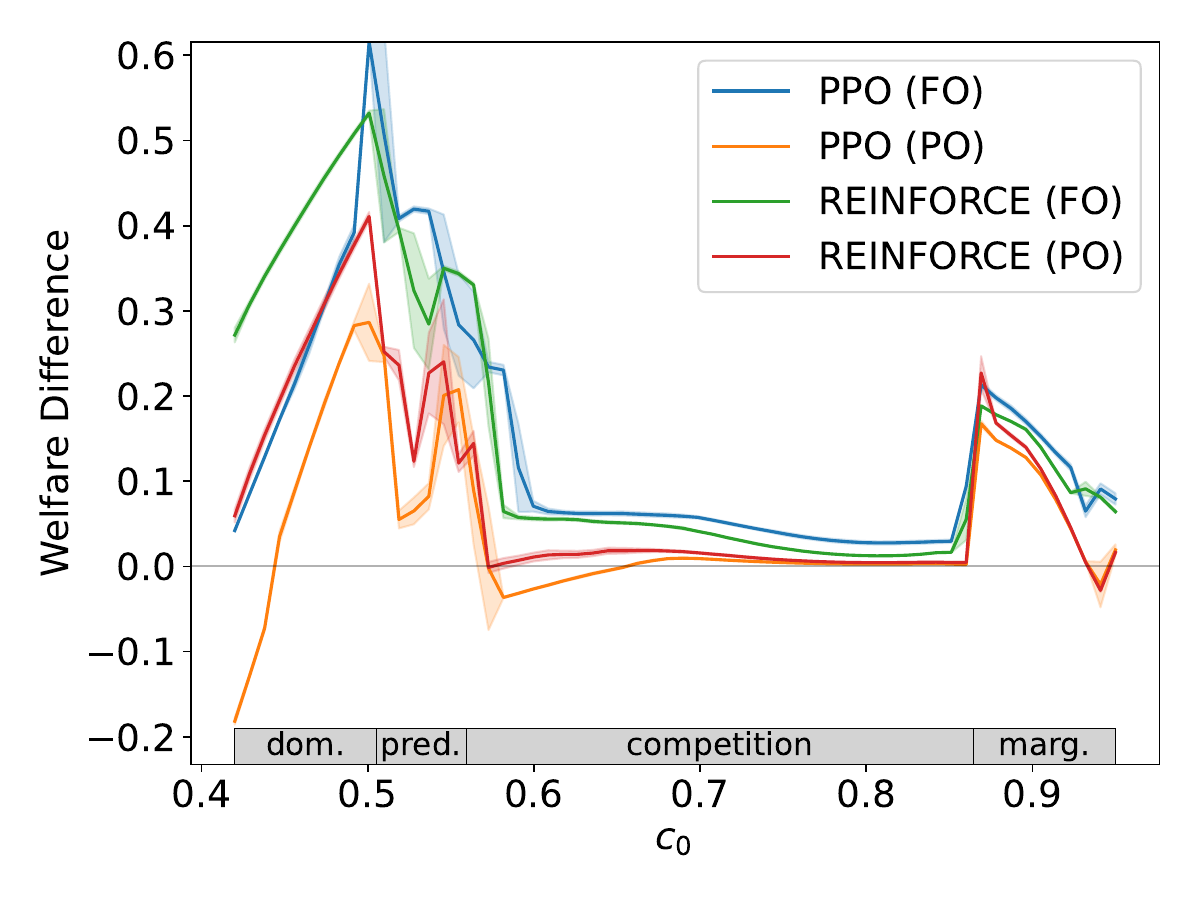}
        \caption{Welfare difference}
        \label{fig:total-welfare-difference}
    \end{subfigure}
    \hfill
    \caption{The producer surplus, consumer surplus, and overall welfare ($\Delta W^{\pi}$) differences for a learned strategy profile $\pi$ and the analytical equilibrium strategies $\pi^*$ without dropout under different costs $c_0$, information structures, and algorithms. The bold line represents the mean and the shaded area the standard deviation over five seeds. The bottom bar indicates the regime, determined by a majority vote over all algorithms, information settings, and random seeds.}
    \label{fig:welfare_differences_dropout_and_analytical}
\end{figure*}

Varying agent~$0$'s unit cost $c_0$ determines its competitive position, resulting in four distinct market regimes: \emph{dominance}, \emph{predation}, \emph{competition}, and \emph{marginalization}. In the \emph{dominance} regime, agent~$0$ leverages its significant cost advantage to eliminate both competitors. Under \emph{predation}, agent~$0$ pushes out one rival and shares the market with the other. As its cost advantage decreases, all agents remain active, producing stable \emph{competition}. Finally, when agent~$0$ is severely disadvantaged, it is driven out by its rivals, defining the \emph{marginalization} regime. These regimes are marked in Figures~\ref{fig:predatory_incentives} and \ref{fig:welfare_differences_dropout_and_analytical} and constitute the main tipping points in behavior.

\begin{table}[b]
\centering
\begin{tabular}{ll|ll|ll|}
\cline{3-6}
 & & \multicolumn{2}{c|}{$ 0.42 \leq c_0 < 0.685 $} & \multicolumn{2}{c|}{$ 0.685 \leq c_0 \leq 0.95 $} \\ \cline{3-6}
\multicolumn{1}{c}{} & & \multicolumn{1}{l|}{$\max_{c_0}\mathcal{L}_{\text{bf}}$} & $\max_{c_0}\mathcal{L}_{\text{bf, norm}}$ & \multicolumn{1}{r|}{$\max_{c_0}\mathcal{L}_{\text{bf}}$} & $\max_{c_0}\mathcal{L}_{\text{bf, norm}}$ \\ \hline
\multicolumn{1}{|l|}{\multirow{2}{*}{PPO (FO)}} & Agent 0 & \multicolumn{1}{l|}{0.032} & 0.048 & \multicolumn{1}{l|}{0.001} & \textcolor{gray}{1.000} \\ \cline{2-6}
\multicolumn{1}{|l|}{} & Agents 1 \& 2 & \multicolumn{1}{l|}{0.010} & \textcolor{gray}{0.496} & \multicolumn{1}{l|}{0.007} & 0.143 \\ \hline
\multicolumn{1}{|l|}{\multirow{2}{*}{PPO (PO)}} & Agent 0 & \multicolumn{1}{l|}{0.019} & 0.050 & \multicolumn{1}{l|}{0.000} & \textcolor{gray}{1.000} \\ \cline{2-6}
\multicolumn{1}{|l|}{} & Agents 1 \& 2 & \multicolumn{1}{l|}{0.009} & \textcolor{gray}{0.477} & \multicolumn{1}{l|}{0.007} & 0.101 \\ \hline
\multicolumn{1}{|l|}{\multirow{2}{*}{REINFORCE (FO)}} & Agent 0 & \multicolumn{1}{l|}{0.021} & 0.046 & \multicolumn{1}{l|}{0.000} & \textcolor{gray}{1.000} \\ \cline{2-6}
\multicolumn{1}{|l|}{} & Agents 1 \& 2 & \multicolumn{1}{l|}{0.008} & \textcolor{gray}{0.440} & \multicolumn{1}{l|}{0.003} & 0.093 \\ \hline
\multicolumn{1}{|l|}{\multirow{2}{*}{REINFORCE (PO)}} & Agent 0 & \multicolumn{1}{l|}{0.021} & 0.045 & \multicolumn{1}{l|}{0.001} & \textcolor{gray}{1.000} \\ \cline{2-6}
\multicolumn{1}{|l|}{} & Agents 1 \& 2 & \multicolumn{1}{l|}{0.007} & \textcolor{gray}{0.411} & \multicolumn{1}{l|}{0.007} & 0.080 \\ \hline
\end{tabular}

\caption{The maximum of the brute force ($\mathcal{L}_{\text{bf}}$) and normalized brute-force ($\mathcal{L}_{\text{bf, norm}}$) losses for the unit cost vector $[c_0,0.8,0.8]$ over all random seed, algorithms, and information settings (FO: Fully observable, PO: Partially observable). Agent~$0$ is reported separately from agents~$1$ and~$2$ because only its unit cost $c_0$ is varied, leading to asymmetric payoffs. Two cost regimes are distinguished to highlight a normalization artifact: When $c_0$ is very low or very high, agent~$0$'s or agent $1$ or $2$'s best-response utility approaches zero, causing even minor absolute deviations (e.g. $< 0.001$) to inflate the normalized loss $\mathcal{L}_{\text{bf,norm}}$ close to $1$. This inflated value does not reflect poor convergence but rather a diminishing denominator. We therefore gray out such values.}
\label{tab:empirical_results:extreme-values-losses}
\end{table}

\textbf{Emergence of predatory behavior}: \Cref{fig:strategy_profiles_large_advantage} shows a strategy profile where agent~$0$ learned predatory pricing, leveraging its significant competitive advantage. Initially, agent~$0$ sets prices close to its production cost, sacrificing short-term profits to push agent~$2$ out by the third round. Subsequently, agents~$0$ and~$1$ raise their prices in the duopoly that follows. This predatory pricing differs substantially from the analytical equilibrium without dropouts, in which agent~$0$ gradually increases prices and agents~$1$ and~$2$ price symmetrically and decrease slightly over time.

\Cref{fig:predatory_incentives} illustrates how predatory incentives depend on agent~$0$'s cost $c_0$, marking the regimes of \emph{dominance}, \emph{predation}, \emph{competition}, and \emph{marginalization}. During \emph{dominance}, agent~$0$ has a strong positive predatory incentive, reflecting significant profitability from monopolizing the market. Predatory incentives decline sharply but remain positive in the \emph{predation} regime, as agent~$0$ benefits by forcing one competitor out. Agents~$1$ and~$2$ also exhibit positive incentives here, as one survives and profits from increased market share. During \emph{competition}, no agents exit, resulting in zero predatory incentives. In the \emph{marginalization} regime, agents~$1$ and~$2$ show increased incentives, aggressively pushing the disadvantaged agent~$0$ from the market. 

Overall, these findings demonstrate that predatory pricing is rational, consistently emerges in equilibrium, and can be robustly learned through independent reinforcement learning algorithms.

\textbf{Welfare effects of predation}: Having established the emergence of predatory behavior, we now assess its welfare implications by comparing the learned strategies (with dropouts) against the analytical equilibrium strategies (without dropouts). The effects of predation on welfare are disputed, as some scholars argue predation reduces consumer welfare by eliminating competition, while others suggest short recoupment phases or uncertain exits may sometimes benefit welfare.

\Cref{fig:welfare_differences_dropout_and_analytical} summarizes the welfare differences in terms of producer surplus ($\Delta PS^{\pi}$), consumer surplus ($\Delta CS^{\pi}$), and total welfare ($\Delta W^{\pi}$). Producer surplus differences largely mirror the predatory incentives: substantial surplus during \emph{dominance}, moderate but positive surplus in \emph{predation}, minimal surplus in \emph{competition}, and an initially sharp increase followed by a decline in \emph{marginalization}. This decrease at high costs occurs because agent~$0$ becomes too uncompetitive to influence the market significantly even when remaining active in the analytical benchmark.

Consumer surplus differences in \cref{fig:consumer-welfare-difference} show distinct patterns. During \emph{competition}, differences remain small. Entering \emph{marginalization}, a notable initial increase occurs due to aggressive price cutting by agents~$1$ and~$2$ to eliminate agent~$0$, but this advantage diminishes as cost differences widen, the sacrifice phase becomes less costly, and the recoupment phase becomes more dominant. A similar effect arises entering \emph{predation}, reflecting high initial sacrifice costs. Another sharp increase occurs as \emph{dominance} begins, followed by a gradual decrease as agent~$0$ leverages its monopoly power earlier and more effectively.

The total welfare difference in \cref{fig:total-welfare-difference} closely follows the consumer surplus pattern. Interestingly, predation-driven exits sometimes enhance overall welfare, especially when inefficient firms exit and aggressive initial price cuts outweigh later price increases. These results indicate that reduced competition can, under certain conditions, lead to better welfare outcomes, challenging traditional antitrust perspectives focused strictly on maximizing competition.

Finally, we observe no significant differences between the fully observable and partially observable settings. Both yield identical market regimes and very similar welfare outcomes, confirming that predatory pricing dynamics primarily depend on timing strategies rather than the granularity of demand information.

These nuanced welfare effects and intricate patterns highlight the importance of using finite-horizon, continuous-action models, which uniquely capture critical timing and trade-off dynamics inaccessible to infinite-horizon or coarser discretized models.

\section{Conclusion}

We analyze predatory pricing behavior in a dynamic oligopoly model extending the seminal framework introduced by~\citet{seltenSpieltheoretischeBehandlungOligopolmodells1965}. By integrating deep reinforcement learning techniques with numerical equilibrium verification, we successfully identify and confirm approximate Nash equilibria that capture realistic predatory strategies. Our finite-horizon model with continuous price-setting addresses previously unresolved questions, allowing us to rigorously analyze the timing of predatory actions and the complex trade-offs between short-term sacrifices and subsequent recoupment. Our results demonstrate that predatory behavior is not only rational and emerges robustly in equilibrium, but also can yield counterintuitive welfare benefits under certain conditions. Specifically, short-term aggressive pricing combined with the removal of inefficient competitors may improve overall market efficiency. These findings challenge conventional antitrust wisdom, underscoring the importance of nuanced analyses that account for timing, cost structures, and competitive dynamics in evaluating market regulation and policy.

\bibliographystyle{unsrtnat}
\bibliography{references}  

\begin{thebibliography}{46}
\providecommand{\natexlab}[1]{#1}
\providecommand{\url}[1]{\texttt{#1}}
\expandafter\ifx\csname urlstyle\endcsname\relax
  \providecommand{\doi}[1]{doi: #1}\else
  \providecommand{\doi}{doi: \begingroup \urlstyle{rm}\Url}\fi

\bibitem[Gates et~al.(1995)Gates, Milgrom, and
  Roberts]{gatesDeterringPredationTelecommunications1995}
Susan Gates, Paul Milgrom, and John Roberts.
\newblock Deterring predation in telecommunications: {{Are}} line-of-business
  restraints needed?
\newblock \emph{Managerial and Decision Economics}, 16\penalty0 (4):\penalty0
  427--438, 1995.

\bibitem[DiLorenzo(1992)]{dilorenzo1992myth}
Thomas~J DiLorenzo.
\newblock \emph{The myth of predatory pricing}.
\newblock Cato Institute, 1992.

\bibitem[May(1994)]{may1994brooke}
Keith~Allen May.
\newblock Brooke group ltd. v. brown \& williamson tobacco corp.: A victory for
  consumer welfare under the robinson-patman act.
\newblock \emph{U. Rich. L. Rev.}, 28:\penalty0 507, 1994.

\bibitem[Bolton et~al.(1999)Bolton, Brodley, and Riordan]{bolton1999predatory}
Patrick Bolton, Joseph~F Brodley, and Michael~H Riordan.
\newblock Predatory pricing: Strategic theory and legal policy.
\newblock \emph{Geo. LJ}, 88:\penalty0 2239, 1999.

\bibitem[Leslie(2023)]{leslie2023predatory}
Christopher~R Leslie.
\newblock Predatory pricing algorithms.
\newblock \emph{NYUL Rev.}, 98:\penalty0 49, 2023.

\bibitem[Cheng and Nowag(2023)]{cheng2023algorithmic}
Thomas~K Cheng and Julian Nowag.
\newblock Algorithmic predation and exclusion.
\newblock \emph{U. Pa. J. Bus. L.}, 25:\penalty0 41, 2023.

\bibitem[Bichler et~al.(2025)Bichler, Durmann, and
  Oberlechner]{bichler2014business}
Martin Bichler, Julius Durmann, and Matthias Oberlechner.
\newblock Algorithmic pricing and algorithmic collusion.
\newblock \emph{Business \& Information Systems Engineering}, to appear, 2025.

\bibitem[Deng et~al.(2024)Deng, Schiffer, and Bichler]{deng2024existence}
Shidi Deng, Maximilian Schiffer, and Martin Bichler.
\newblock On the existence of algorithmic collusion in dynamic pricing with
  deep reinforcement learning.
\newblock \emph{Conference on Wirtschaftsinformatik}, 2024.

\bibitem[Sanders et~al.(2018)Sanders, Farmer, and Galla]{sanders2018prevalence}
James~BT Sanders, J~Doyne Farmer, and Tobias Galla.
\newblock The prevalence of chaotic dynamics in games with many players.
\newblock \emph{Scientific {R}eports}, 8\penalty0 (1):\penalty0 1--13, 2018.

\bibitem[Pieroth et~al.(2025)Pieroth, Kohring, and Bichler]{pieroth2025}
Fabian~Raoul Pieroth, Nils Kohring, and Martin Bichler.
\newblock Deep reinforcement learning for equilibrium computation in
  multi-stage auctions and contests.
\newblock \emph{Management Science}, 2025.

\bibitem[Milgrom(1990)]{milgromNewTheoriesPredatory1990}
Paul Milgrom.
\newblock New theories of predatory pricing.
\newblock \emph{Industrial structure in the new industrial economics}, 1990.

\bibitem[Telser(1966)]{telserCutthroatCompetitionLong1966}
L.~G. Telser.
\newblock Cutthroat {{Competition}} and the {{Long Purse}}.
\newblock \emph{The Journal of Law \& Economics}, 9:\penalty0 259--277, 1966.

\bibitem[Nash(1950)]{nash1950equilibrium}
John~Fs Nash.
\newblock Equilibrium points in n-person games.
\newblock \emph{Proceedings of the national academy of sciences}, 36\penalty0
  (1):\penalty0 48--49, 1950.

\bibitem[Maskin and Tirole(1988)]{maskin1988theory}
Eric Maskin and Jean Tirole.
\newblock A theory of dynamic oligopoly, ii: Price competition, kinked demand
  curves, and edgeworth cycles.
\newblock \emph{Econometrica: Journal of the Econometric Society}, pages
  571--599, 1988.

\bibitem[Cabral and Riordan(1994)]{cabralLearningCurveMarket1994}
Luis Cabral and Michael Riordan.
\newblock The {{Learning Curve}}, {{Market Dominance}}, and {{Predatory
  Pricing}}.
\newblock \emph{Econometrica}, 62\penalty0 (5):\penalty0 1115--40, 1994.

\bibitem[Besanko et~al.(2011)Besanko, Doraszelski, and
  Kryukov]{besankoEconomicsPredationWhat}
David Besanko, Ulrich Doraszelski, and Yaroslav Kryukov.
\newblock The {{Economics}} of {{Predation}}: {{What Drives Pricing When
  There}} is {{Learning-by-Doing}}?
\newblock \emph{GSIA Working Papers}, \penalty0 (E8), 2011.

\bibitem[Rey et~al.(2022)Rey, Spiegel, and Stahl]{reyDynamicModelPredation2022}
Patrick Rey, Yossi Spiegel, and Konrad~O. Stahl.
\newblock A {{Dynamic Model}} of {{Predation}}.
\newblock 2022.

\bibitem[Fudenberg and Kreps(1993)]{fudenberg1993learning}
Drew Fudenberg and David~M Kreps.
\newblock Learning mixed equilibria.
\newblock \emph{Games and Economic Behavior}, 5\penalty0 (3):\penalty0
  320--367, 1993.

\bibitem[Pakes and McGuire(1992)]{pakes1992computing}
Ariel Pakes and Paul McGuire.
\newblock Computing markov perfect nash equilibria: Numerical implications of a
  dynamic differentiated product model, 1992.

\bibitem[Schulman et~al.(2017)Schulman, Wolski, Dhariwal, Radford, and
  Klimov]{DBLP:journals/corr/SchulmanWDRK17}
John Schulman, Filip Wolski, Prafulla Dhariwal, Alec Radford, and Oleg Klimov.
\newblock Proximal policy optimization algorithms.
\newblock \emph{CoRR}, abs/1707.06347, 2017.

\bibitem[Selten(1965)]{seltenSpieltheoretischeBehandlungOligopolmodells1965}
Reinhard Selten.
\newblock Spieltheoretische {{Behandlung Eines Oligopolmodells Mit
  Nachfragetr{\"a}gheit}}: {{Teil I}}: {{Bestimmung Des Dynamischen
  Preisgleichgewichts}}.
\newblock \emph{Zeitschrift f{\"u}r die gesamte Staatswissenschaft / Journal of
  Institutional and Theoretical Economics}, 121\penalty0 (2):\penalty0
  301--324, 1965.

\bibitem[Bertrand(1883)]{bertrandTheorieMathematiqueRichesse1883}
Jean Bertrand.
\newblock Th{\'e}orie math{\'e}matique de la richesse sociale.
\newblock \emph{Journal des Savants}, \penalty0 (68):\penalty0 499--508, 1883.

\bibitem[Bylka et~al.(2000)Bylka, Ambroszkiewicz, and
  Komar]{bylkaDiscreteTimeDynamic2000}
Stanis{\l}aw Bylka, Stanis{\l}aw Ambroszkiewicz, and Jan Komar.
\newblock Discrete time dynamic game model for price competition in an
  oligopoly.
\newblock \emph{Annals of Operations Research}, 97\penalty0 (1):\penalty0
  69--89, 2000.

\bibitem[Assad et~al.(2020)Assad, Clark, Ershov, and Xu]{gasoline}
Stephanie Assad, Robert Clark, Daniel Ershov, and Lei Xu.
\newblock Algorithmic pricing and competition: Empirical evidence from the
  german retail gasoline market.
\newblock CESifo Working Paper 8521, CESifo, 2020.
\newblock Available at SSRN: \url{https://ssrn.com/abstract=3682021} or
  \url{http://dx.doi.org/10.2139/ssrn.3682021}.

\bibitem[Madhavan(2000)]{financial_markets}
Ananth Madhavan.
\newblock Market microstructure: A survey.
\newblock \emph{Journal of Financial Markets}, 3\penalty0 (3):\penalty0
  205--258, 2000.
\newblock ISSN 1386-4181.
\newblock \doi{https://doi.org/10.1016/S1386-4181(00)00007-0}.
\newblock URL
  \url{https://www.sciencedirect.com/science/article/pii/S1386418100000070}.

\bibitem[Escobari and Lee(2014)]{airline_demand_uncertainty}
Diego Escobari and Jim Lee.
\newblock Demand uncertainty and capacity utilization in airlines.
\newblock \emph{Empirical Economics}, 47, 08 2014.
\newblock \doi{10.1007/s00181-013-0725-2}.

\bibitem[Fudenberg and Tirole(2013)]{fudenberg2013dynamic}
Drew Fudenberg and Jean Tirole.
\newblock \emph{Dynamic models of oligopoly}.
\newblock Routledge, 2013.

\bibitem[Gerpott and Berends(2022)]{gerpottCompetitivePricingOnline2022}
Torsten~J. Gerpott and Jan Berends.
\newblock Competitive pricing on online markets: A literature review.
\newblock \emph{Journal of Revenue and Pricing Management}, 21\penalty0
  (6):\penalty0 596--622, December 2022.

\bibitem[Phlips and Richard(1989)]{phlips1989dynamic}
Louis Phlips and Jean-Francois Richard.
\newblock A dynamic oligopoly model with demand inertia and inventories.
\newblock \emph{Mathematical Social Sciences}, 18\penalty0 (1):\penalty0 1--32,
  1989.

\bibitem[Farrell and Shapiro(1988)]{farrell1988dynamic}
Joseph Farrell and Carl Shapiro.
\newblock Dynamic competition with switching costs.
\newblock \emph{The RAND Journal of Economics}, pages 123--137, 1988.

\bibitem[Bayer and Chan(2007)]{bayer2007network}
Ralph-C Bayer and Mickey Chan.
\newblock Network externalities, demand inertia and dynamic pricing in an
  experimental oligopoly market.
\newblock \emph{Economic Record}, 83\penalty0 (263):\penalty0 405--415, 2007.

\bibitem[Besanko et~al.(2014)Besanko, Doraszelski, and
  Kryukov]{besanko2014economics}
David Besanko, Ulrich Doraszelski, and Yaroslav Kryukov.
\newblock The economics of predation: What drives pricing when there is
  learning-by-doing?
\newblock \emph{American Economic Review}, 104\penalty0 (3):\penalty0 868--897,
  2014.

\bibitem[Fudenberg and Levine(1999)]{fudenbergTheoryLearningGames1999}
Drew Fudenberg and David~K. Levine.
\newblock \emph{The Theory of Learning in Games}, volume~2 of \emph{{{MIT
  Press}} Series on Economic Learning and Social Evolution}.
\newblock MIT Press, Cambridge, 2. edition, 1999.

\bibitem[Milionis et~al.(2023)Milionis, Papadimitriou, Piliouras, and
  Spendlove]{milionisImpossibilityTheoremGame2023}
Jason Milionis, Christos Papadimitriou, Georgios Piliouras, and Kelly
  Spendlove.
\newblock An impossibility theorem in game dynamics.
\newblock \emph{Proceedings of the National Academy of Sciences}, 120\penalty0
  (41):\penalty0 e2305349120, October 2023.

\bibitem[Mazumdar et~al.(2020)Mazumdar, Ratliff, Jordan, and
  Sastry]{mazumdarPolicyGradientAlgorithmsHave2020}
Eric Mazumdar, Lillian~J. Ratliff, Michael~I. Jordan, and S.~Shankar Sastry.
\newblock Policy-{{Gradient Algorithms Have No Guarantees}} of {{Convergence}}
  in {{Linear Quadratic Games}}.
\newblock In \emph{International {{Conference}} on {{Autonomous Agents}} and
  {{Multi Agent Systems}} ({{AAMAS}})}, {{AAMAS}} '20, pages 860--868,
  Richland, SC, May 2020. {International Foundation for Autonomous Agents and
  Multiagent Systems}.

\bibitem[Daskalakis et~al.(2010)Daskalakis, Frongillo, Papadimitriou,
  Pierrakos, and Valiant]{daskalakisLearningAlgorithmsNash2010}
Constantinos Daskalakis, Rafael Frongillo, Christos~H Papadimitriou, George
  Pierrakos, and Gregory Valiant.
\newblock On learning algorithms for {{Nash}} equilibria.
\newblock In \emph{International Symposium on Algorithmic Game Theory}, pages
  114--125. Springer, 2010.

\bibitem[Bichler et~al.(2023)Bichler, Lunowa, Oberlechner, Pieroth, and
  Wohlmuth]{bichlerConvergenceLearningAlgorithms2023}
Martin Bichler, Stephan~B. Lunowa, Matthias Oberlechner, Fabian~R. Pieroth, and
  Barbara Wohlmuth.
\newblock On the {{Convergence}} of {{Learning Algorithms}} in {{Bayesian
  Auction Games}}, November 2023.

\bibitem[{\c S}eref~Ahunbay and
  Bichler(2024)]{serefahunbayUniquenessBayesianCoarse2024}
Mete {\c S}eref~Ahunbay and Martin Bichler.
\newblock On the {{Uniqueness}} of {{Bayesian Coarse Correlated Equilibria}} in
  {{Standard First-Price}} and {{All-Pay Auctions}}, January 2024.

\bibitem[Albrecht et~al.(2024)Albrecht, Christianos, and Sch\"afer]{marl-book}
Stefano~V. Albrecht, Filippos Christianos, and Lukas Sch\"afer.
\newblock \emph{Multi-Agent Reinforcement Learning: Foundations and Modern
  Approaches}.
\newblock MIT Press, 2024.
\newblock URL \url{https://www.marl-book.com}.

\bibitem[{van de Geer} et~al.(2019){van de Geer}, {den Boer}, Bayliss, Currie,
  Ellina, Esders, Haensel, Lei, Maclean, {Martinez-Sykora}, Riseth,
  {\O}degaard, and Zachariades]{vandegeerDynamicPricingLearning2019}
Ruben {van de Geer}, Arnoud~V. {den Boer}, Christopher Bayliss, Christine S.~M.
  Currie, Andria Ellina, Malte Esders, Alwin Haensel, Xiao Lei, Kyle D.~S.
  Maclean, Antonio {Martinez-Sykora}, Asbj{\o}rn~Nilsen Riseth, Fredrik
  {\O}degaard, and Simos Zachariades.
\newblock Dynamic pricing and learning with competition: Insights from the
  dynamic pricing challenge at the 2017 {{INFORMS RM}} \& pricing conference.
\newblock \emph{Journal of Revenue and Pricing Management}, 18\penalty0
  (3):\penalty0 185--203, June 2019.

\bibitem[Courty(2000)]{a06c4670-206d-3c08-987c-97337c3ffec5}
Pascal Courty.
\newblock An economic guide to ticket pricing in the entertainment industry.
\newblock \emph{Recherches Économiques de Louvain / Louvain Economic Review},
  66\penalty0 (2):\penalty0 167--192, 2000.
\newblock ISSN 07704518, 17821495.
\newblock URL \url{http://www.jstor.org/stable/40724285}.

\bibitem[Sutton and Barto(2018)]{SuttonAndBarto2018}
Richard~S Sutton and Andrew~G Barto.
\newblock \emph{Reinforcement Learning: {{An}} Introduction}.
\newblock A Bradford Book, Cambridge, Massachusetts, 2 edition, 2018.

\bibitem[Raffin et~al.(2021)Raffin, Hill, Gleave, Kanervisto, Ernestus, and
  Dormann]{raffinStableBaselines3ReliableReinforcement2021}
Antonin Raffin, Ashley Hill, Adam Gleave, Anssi Kanervisto, Maximilian
  Ernestus, and Noah Dormann.
\newblock Stable-{{Baselines3}}: {{Reliable Reinforcement Learning
  Implementations}}.
\newblock \emph{Journal of Machine Learning Research}, 22\penalty0
  (268):\penalty0 1--8, 2021.

\bibitem[Ordover and Willig(1981)]{ordoverEconomicDefinitionPredation1981}
Janusz~A. Ordover and Robert~D. Willig.
\newblock An {{Economic Definition}} of {{Predation}}: {{Pricing}} and
  {{Product Innovation}}.
\newblock \emph{The Yale Law Journal}, 91\penalty0 (1):\penalty0 8--53, 1981.

\bibitem[Belleflamme and
  Peitz(2010)]{belleflammeIndustrialOrganizationMarkets2010}
Paul Belleflamme and Martin Peitz.
\newblock \emph{Industrial {{Organization}}: {{Markets}} and {{Strategies}}}.
\newblock Cambridge University Press, 1 edition, January 2010.

\bibitem[Petrazzini and
  Antonelo(2021)]{petrazzini2021proximalpolicyoptimizationcontinuous}
Irving G.~B. Petrazzini and Eric~A. Antonelo.
\newblock Proximal policy optimization with continuous bounded action space via
  the beta distribution, 2021.
\newblock URL \url{https://arxiv.org/abs/2111.02202}.

\end{thebibliography}

\begin{acronym}
    \itemsep-.25\baselineskip
    \acro{RL}[RL]{reinforcement learning}
    \acro{MDP}[MDP]{Markov decision process}
    \acro{POMDP}[POMDP]{partially observable Markov decision process}
    \acro{POMG}[POMG]{partially observable Markov game}
    \acroplural{MDP}[MDPs]{Markov decision processes}
    \acro{NE}[NE]{Nash equilibrium}
    \acroplural{NE}[NE]{Nash equilibria}
    \acroplural{BNE}[BNE]{Bayes Nash equilibria}
    \acro{MARL}[MARL]{multi-agent reinforcement learning}
    \acro{PPO}[PPO]{proximal policy optimization}
\end{acronym}






\end{document}